\documentclass[final,5p,times,twocolumn]{elsarticle}

 \usepackage{graphicx}
 \usepackage{epsfig}
\usepackage{amssymb}
\usepackage{amssymb}
\usepackage{amsthm, amsmath, amssymb}

\usepackage[dvips]{color} 

\newcommand{\ncd}{\newcommand}
	\ncd{\mrm}    {\mathrm}
	\ncd{\beq} {\begin{equation}}
	\ncd{\eeq} {\end{equation}}

	\def\d{{\rm d}}

\begin{document}

\begin{frontmatter}



\title{A static axisymmetric  exact solution of $f(R)$-gravity }
 
\author[rvt,focal]{Antonio C. Guti\'errez-Pi\~{n}eres\corref{cor1}}
\ead{acgutierrez@correo.nucleares.unam.mx}

\author[focal]{C\'esar. S. L\'opez-Monsalvo}
\ead{cesar.slm@correo.nucleares.unam.mx}

\cortext[cor1]{Corresponding author: Faculty of Basic Sciences,
Universidad Tecnol\'ogica de Bol\'ivar, 
 CO 131001, Cartagena de Indias, Colombia}

\address[rvt]{Facultad de Ciencias B\'asicas, Universidad Tecnol\'ogica
 de Bol\'ivar, CO 131001 Cartagena de Indias, Colombia}

\address[focal]{Instituto de Ciencias Nucleares, Universidad Nacional 
Aut\'onoma de M\'exico, A.P. 70-543, 04510 México D.F., Mexico}

\begin{abstract}
We present an exact, axially symmetric, static, vacuum  solution for $f(R)$ gravity in Weyl's canonical coordinates.  We obtain a general explicit expression for the
dependence of $df(R)/dR$ upon the $r$ and $z$ coordinates and then the corresponding explicit form of $f(R)$, which must be consistent with the field equations.  We analyze in detail the modified Schwarzschild solution in prolate spheroidal coordinates. Finally, we study the curvature invariants and show that, in the case of $f(R)\neq R$, this solution corresponds to a naked singularity.
\end{abstract}


\end{frontmatter}

\section{Introduction}
\label{sec:intro}

In recent years, $f(R)$ theories of gravity have gained much attention as  promising 
candidates to overcome the issues posted by dark energy in the standard cosmological model 
(c.f. reference \cite{RevModPhys.82.451, 2011PhR...505...59N} 
for a recent review). There has been a stimulating debate in their study, leading us to a number 
of interesting results, particularly in the context of exact solutions.

Due to the highly non-linear nature of the $f(R)$-field equations, finding exact solutions
 is indeed a difficult task. Following astrophysical motivations, solutions with 
spherical symmetry have been the most widely studied \cite{PhysRevD.74.064022}. However, 
there has also been a growing interest in finding exact solutions with cylindrical symmetry \cite{Azadi2008210}.
 To the best of the authors' knowledge (except for \cite{0264-9381-27-16-165008,
Cembranos:2011sr}), there is no fully integrated, explicit and exact axially symmetric solution of $f(R)$-gravity.

In this paper, we consider static vacuum solutions of $f(R)$ theories in
Weyl coordinates. In particular, we obtain the explicit dependence of $df(R)/dR$ upon the 
coordinates $\rho$ and $z$. This in turn, allows us to get the corresponding explicit form of $f(R)$. 
Finally, we analyze in detail the solutions of the modified field
equations corresponding to the Schwarzschild solution in cylindrical coordinates.

\section{\label{sec:FE} $f(R)$ Field equations for a static  axially symmetric space-time}

The $f(R)$ action is given by 
\begin{eqnarray}
 S=\int{\left(\frac{1}{16\pi G}f(R) + {\cal L}_{\text
m}\right)\sqrt{-g}\ \d^4 x},\label{eq:action}
\end{eqnarray}
where $G$ is the gravitational  constant, $R$ is  the curvature scalar and
${\cal L}_{\text
m}$ is  the matter Lagrangian.
The   field equation  resulting from  this  action are 
 \begin{eqnarray}
 G_{ab} &=& R_{ab} - \frac{1}{2}Rg_{ab} = 8\pi G(\tilde T_{ab}^{\text{g}}
+ \tilde T_{ab}^{\text m}),\label{eq:fe}
 \end{eqnarray}
where  the `gravitational' stress-energy tensor is 
\begin{eqnarray}
 8 \pi G \tilde T_{ab}^{\text g} = T_{ab}^{\text g},
\end{eqnarray} 
and
\begin{align*}
T_{ab}^{\text g}=
 \frac{1}{F_R}\left[\frac{g_{ab}}{2}\left(f(R) - RF_R\right) 
 +  \nabla_c \nabla_d F_R \left(\delta_{a}^{\ c}
 \delta_{b}^{\ d} - g_{ab} g^{cd}\right)\right].
\end{align*}
Here $F_R\equiv df(R)/dR$ and $\tilde T_{ab}^{\text m}\equiv T_{ab}^{\text m}/F_R$, where
$T_{ab}^{\text m}$ is the stress-energy tensor obtained from the matter Lagrangian ${\cal
L}_{\text m}$ in the  action (\ref{eq:action}) .

Equivalently, we can write (\ref{eq:fe}) in the  form
\begin{equation}
 F_R R_{ab} - \frac{1}{2}f(R)g_{ab} - \nabla_{a}\nabla_{b}F_R
+ g_{ab} \Box F_R = 8\pi G  T_{ab}^{\text m}. \label{eq:MFE}
\end{equation}
Taking the trace of this expression we obtain the relation between $f(R)$ and its derivative 
$F_R$ 
\begin{equation}
 F_R R - 2f(R) + 3\Box F_R = 8\pi G T^{\text m}. \label{eq:contrac}
\end{equation}

We are interested in the  static axially symmetric solutions of (\ref{eq:MFE}). To this end, let us consider the Weyl-
Lewis-Papapetrou metric in cylindrical coordinates is  \cite{KSMH}
\begin{equation}
 ds^2= -e^{2\phi}dt^2 + e^{-2\phi}[\rho^2d\varphi^2 + e^{2\lambda}
 (d\rho^2 + dz^2)]\label{eq:metric},
\end{equation}
where $\phi$ and $\lambda$ are continuous functions of $\rho$ and $z$.
Using the trace equation (\ref{eq:contrac}), the modified Einstein field equations 
(\ref{eq:MFE})
become
\begin{equation}
 F_R R_{ab} - \nabla_{a}\nabla_{b}F_R - 8\pi GT_{ab}^{\text m}
= {g_{ab}}B, \label{eq:modified}
\end{equation}
where $B= \frac{1}{4}(F_RR - \Box F_R -8\pi G T^{\text m})$. The non-zero components of  the Ricci tensor are 
\begin{subequations}\begin{eqnarray}
&& R_{00} = e^{4\phi -2\lambda}\nabla^2\phi, \\
&&R_{11} = \rho^2e^{-2\lambda}\nabla^2\phi,\\
&&R_{22} = -\nabla^2\lambda + \nabla^2\phi + \frac{2}{\rho}\lambda_{,\rho}
- 2(\phi_{,\rho})^2,\\
&&R_{33} = -\nabla^2\lambda + \nabla^2\phi 
- 2\phi_{,z}^2,\\
&&R_{23} = \frac{1}{\rho}\lambda_{,z}- 2\phi_{,\rho}\phi_{,z},
\end{eqnarray}\label{eq:ricci}\end{subequations}
while  a straightforward computation  of the curvature
scalar yields
\begin{equation}
 R=2e^{2\phi - 2\lambda}(-\nabla^2\lambda + \nabla^2\phi +
\frac{1}{\rho}\lambda_{,\rho} - \phi_{,\rho}^2 - \phi_{,z}^2),
\end{equation}
where $\nabla^2$ is the usual  Laplace operator in cylindrical coordinates. 

From (\ref{eq:modified}) and (\ref{eq:ricci}) we obtain the following system of equations:
\begin{subequations}
\begin{align}
\frac{1}{g_{00}} \left[ F_RR_{00} - 8\pi GT_{00}^{\text m}\right] & = B,\\
\frac{1}{g_{11}} \left[ F_RR_{11} - 8\pi GT_{11}^{\text m}\right] & = B,\\
\frac{1}{g_{22}} \left[ F_RR_{22} - 8\pi GT_{22}^{\text m}\right] & = B,\\
\frac{1}{g_{33}} \left[ F_RR_{33} - 8\pi GT_{33}^{\text m}\right] & = B,\\
F_RR_{23} - 8\pi GT_{23}^{\rm m} -  F_{R,23} & = 0.
\end{align}
\end{subequations}

This allows us to write down the independent field equations, i.e.
\begin{subequations}
\begin{align}
\nabla^2\phi  = &\  -\frac{4\pi G}{F_R}e^{2(\lambda - \phi)}[T_{\;0}^{\text m 0} - T_{\;
1}^{\text m 1}],\label{eq:MEEn}\\
\lambda_{,\rho}  = &\  \rho(\phi_{,\rho}^2 - \phi_{,z}^2)+
\frac{4\pi G\rho}{F_R}e^{2(\lambda - \phi)}[T_{\;2}^{\text m 2} - T_{\;3}^{\text m 3}] \nonumber\\
& + \frac{\rho}{2F_R}[F_{R,22} -F_{R,33}],\label{eq:MEE1}\\
\lambda_{,z}  = &\  2\rho\phi_{,\rho}\phi_{,z}+ \frac{8\pi G\rho}{F_R} T_{\;23}^{\text m}
+ 
\frac{\rho F_{R,23}}{F_R}\label{eq:MEE2}.
\end{align}\label{eq:MEE3}
\end{subequations}

To find a general solution to the  above  equations is indeed a difficult task. Nevertheless, in the following sections we will discuss some particular solutions to \eqref{eq:MEEn}.

\section{\label{sec:ES} $f(R)$ vacuum solutions for a static axially symmetric space-time}
For simplicity, we restrict ourselves to the vacuum case, i.e. we make
\begin{subequations}
\begin{eqnarray}
&&\nabla^2\phi =0,\label{eq:laplace}\\
&&\lambda_{,\rho}=\rho(\phi_{,\rho}^2 - \phi_{,z}^2) 
+ \frac{\rho}{2F_R}(F_{R,\rho\rho} -F_{R,zz}),\label{eq:laplacelambda1}\\
&&\lambda_{,z}  = 2\rho\phi_{,\rho}\phi_{,z}
+ \frac{\rho F_{R,\rho z}}{F_R},\label{eq:laplacelambda2}\\
&&\rho F_{R,z}(F_{R,\rho\rho} - F_{R,zz})
+ \rho F_R\nabla^2(F_{R,z})\nonumber \\
&&\qquad\qquad\;\;\; +\;\; F_{R,z\rho}(F_R - 2\rho F_{R,\rho}) = 0,\label{eq:integrab}\\
&&R=-2e^{2\phi - 2\lambda}(\lambda_{,\rho\rho}  + \lambda_{,zz} + 
\phi_{,\rho}^2 + \phi_{,z}^2),\label{eq:R}\\
&&f(R)=\frac{1}{2}F_RR + \frac{3}{2} \Box F_R\label{eq:f(R)}.
\end{eqnarray}\label{eq:MEEvac}
\end{subequations}
One can see that equation (\ref{eq:integrab}) is the integrability condition for $\lambda$. Note that for an arbitrary $f(R)$, the system \eqref{eq:MEEvac} may become inconsistent. Therefore, we look for the class of functions $f(R)$ compatible with \eqref{eq:MEEvac}. It is an easy exercise to prove that 
\begin{equation}
\Box F_R = e^{2(\phi - \lambda)}(F_{R,\rho\rho} + F_{R,zz}).\label{eq:boxF_R}
\end{equation}
Thus, substituting  (\ref{eq:boxF_R}) in (\ref{eq:f(R)}) and using 
(\ref{eq:R}) we obtain
\begin{eqnarray}
f(R)=\frac{1}{2}F_RR\left[1 - \frac{3W(\rho)}{2F_R}\right],
\label{eq:f(R)exp}
\end{eqnarray}
where
\begin{eqnarray}
W(\rho)=\frac{F_{R,\rho\rho} + F_{R,zz}}{\lambda_{,\rho\rho}  + \lambda_{,zz} + 
\phi_{,\rho}^2 + \phi_{,z}^2}.
\end{eqnarray}

In order to obtain some analytical solutions to the system (\ref{eq:MEEvac}), we need to make some 
further simplifying assumptions. First, suppose that it is possible to
write 
\begin{eqnarray}
\label{FR:sep}
 F_R(\rho,z)=U(\rho)V(z).
\end{eqnarray}
Then, substituting back into (\ref{eq:integrab}) we have
\begin{eqnarray}
 \rho^{-1}U^{-2}\left[2\frac{dU}{d\rho}
\left(\rho\frac{dU}{d\rho}-U\right) - 2\rho U\frac{d^2U}{d\rho^2}\right] \nonumber\\=
\left(V\frac{dV}{dz}\right)^{-1}\left[V\frac{d^3V}{dz^3}-\frac{d^2V}{dz^2}\frac{dV}{dz}\right].
\end{eqnarray}
Equating each side  to a separation constant, $l^2$,  we obtain the third order pair of ordinary differential equations
\begin{subequations}
 \begin{eqnarray}
 \rho^{-1}U^{-2}\left[2\frac{dU}{d\rho}\left(\rho\frac{dU}{d\rho}-U\right) 
 - 2\rho U\frac{d^2U}{d\rho^2}\right]=l^2\label{eq:sep1}\\
\left(V\frac{dV}{dz}\right)^{-1}
\left[V\frac{d^3V}{dz^3}-\frac{d^2V}{dz^2}\frac{dV}{dz}\right]=l^2. \label{eq:sep2}
\end{eqnarray}
\end{subequations}

Let us re-write (\ref{eq:sep1}) as
\begin{equation}
 \frac{dM(\rho)}{d\rho} + \frac{M(\rho)}{\rho}=-\frac{l^2}{2}\label{eq:mequation},
\end{equation}
where $M(\rho)=U^{-1}dU/d\rho$. One can solve this immediately to obtain  
\begin{eqnarray}
 M(\rho)=U^{-1}\frac{dU}{d\rho}= \frac{n}{\rho} - \frac{l^2\rho}{4},
\end{eqnarray}
which has the solution
\begin{eqnarray}
U(\rho)=c\rho^ne^{{-l^2\rho^2}/{8}},
\end{eqnarray}
where $c$ and $n$ are integration constants.  

Now, one can easily show that 
\begin{eqnarray}
 V(z)= e^{bz}\label{eq:vsol},
\end{eqnarray}
where $b$ is an arbitrary constant, is solution of (\ref{eq:sep2}) if both, $b$ and $l$,
satisfy the condition
\begin{eqnarray}
 bl^2=0.
\end{eqnarray}
Thus, we have  that some possible solutions for $F_R$ are [c.f. equation \eqref{FR:sep}]
\begin{enumerate}[(\bf i)]
 \item $b=0$ and $l=0$.
In this case 
\begin{eqnarray}
 F_R=c\rho^n.\label{eq:FR1}
\end{eqnarray}
 \item $b\neq 0$ and $l=0$.
In this case 
\begin{eqnarray}
  F_R=c\rho^ne^{bz}.\label{eq:FR2}
\end{eqnarray}
 \item $b=0$ and $l\neq 0$.
In this case 
\begin{eqnarray}
 F_R=c\rho^ne^{-l^2\rho^2/8}.\label{eq:FR3}
\end{eqnarray}
\end{enumerate}
Consequently, by substituting (\ref{eq:FR1}) [or (\ref{eq:FR2})] in
(\ref{eq:f(R)exp}) and using \eqref{eq:MEEvac} we obtain that $f(R)$ must satisfy the consistency condition
\begin{eqnarray}
  f(R)= 2R\frac{df}{dR},
 \end{eqnarray}
whose  solution is simply
\begin{equation}
 f(R)=kR^{1/2},
\end{equation}
where $k$ is an arbitrary constant.

Substituting (\ref{eq:FR2}) in both, equations (\ref{eq:laplacelambda1}) and 
(\ref{eq:laplacelambda2}), we  have
\begin{subequations}\begin{eqnarray}
&&\lambda_{,\rho}=\rho(\phi_{,\rho}^2 - \phi_{,z}^2) 
+ \frac{1}{2\rho}[n(n-1) -b^2\rho^2],\label{eq:lambdarpart}\\
&&\lambda_{,z}  = 2\rho\phi_{,\rho}\phi_{,z}
+ bn,\label{eq:lambdazpart}
\end{eqnarray}\label{eq:lambdaI}\end{subequations}
respectively. Whereas, by taking 
$F_R=c\rho^ne^{-l^2\rho^2/8}$ we obtain from  (\ref{eq:f(R)exp})
\begin{eqnarray}
 f(R)=\frac{1}{2}F_RR\left[1 - 3L(\rho)\right],\label{eq:f(R)3}
\end{eqnarray}
with
\begin{eqnarray}
 L(\rho)=\frac{(l^2\rho^2 - 4n)^2 - 4(l^2\rho^2 + 4n)}
{(3l^2\rho^2 + 4n -4)(l^2\rho^2 - 4n)}.
\end{eqnarray}

Finally, substituting (\ref{eq:FR3}) in (\ref{eq:laplacelambda1}) and 
(\ref{eq:laplacelambda2}), we  have
\begin{subequations}\begin{eqnarray}
\lambda_{,\rho}&=&\rho(\phi_{,\rho}^2 - \phi_{,z}^2) 
\nonumber\\&+& \frac{1}{32\rho}
[(l^2\rho^2-4n)^2 -4(l^2\rho^2 + 4n)],\label{eq:lambdarpart2}\\
\lambda_{,z}  &=& 2\rho\phi_{,\rho}\phi_{,z},\label{eq:lambdazpart2}
\end{eqnarray}\label{eq:lambdaII}\end{subequations}
respectively.


As  we  can see from equations (\ref{eq:lambdaI}) and (\ref{eq:lambdaII}), the function $\lambda$
can be calculated  by means of a line integral. Although $\nabla^2\phi =0$ is a linear
differential equation, the equations for $\lambda$ manifest the non-linearity of the
``modified'' Einstein field equations.
 
The usual Einstein vacuum equations for  the static  axisymmetric spacetime we have 
\begin{subequations}
\begin{eqnarray}
&&\nabla^2{\phi} =0,\label{eq:laplacevacuum}\\
&&\lambda_{,\rho}=\rho(\phi_{,\rho}^2 - \phi_{,z}^2) 
,\label{eq:laplacelambda1vaccum}\\
&&\lambda_{,z}  = 2\rho\phi_{,\rho}\phi_{,z}.\label{eq:laplacelambda2vacuum}
\end{eqnarray}\label{eq:EEvac}
\end{subequations}
The Laplace equation may be solved by using various coordinates in the
Euclidean 3-space and then the function $\lambda$ can be calculated (c.f. chapter 20 in \cite{KSMH}). We observe the following points:
\begin{enumerate}
 \item If  $F_R=c\rho^ne^{bz}$, then we  can obtain a vacuum
static axially symmetric solution $(\phi,\lambda)$ of the vacuum `modified' Einstein
field equations from the vacuum Einstein field equations $({\tilde \phi},
{\tilde \lambda})$ using the  transformation
\begin{subequations}\begin{eqnarray}
 \phi&=&{\tilde \phi}, \\
 \lambda&= &{\tilde \lambda} + \ln\left[k\rho^{n(n-1)/2}\right] - \frac{b^2}{4}\rho^2 +
bnz.
\end{eqnarray}\label{eq:WeylModI}\end{subequations}
\item Similarly, if $ F_R=c\rho^ne^{-l^2\rho^2/8}$ we make
\begin{subequations}\begin{align}
\phi&={\tilde \phi}, \\
\lambda&={\tilde \lambda} \nonumber\\&+
\frac{1}{32}\left\{\frac{l^4\rho^4}{4} - 2(2n+1)l^2\rho^2 + \ln\left[k\rho^{16n(n-1)}
\right]
\right\}.
\end{align}\label{eq:WeylModII}\end{subequations}
In both cases, $k$ is an appropriate constant.
\end{enumerate}
                                                
Thus, one can say that equations (\ref{eq:WeylModI}) and (\ref{eq:WeylModII})
represent a {\it  Weyl class of solutions} in $f(R)$-gravity. Moreover, one can expect
that some properties of the curvature of the seed solution will be inherited to the
modified ones.


\hyphenation{va-cuum}
\section{\label{sec:part} A particular solution for a vacuum static axially symmetric 
space-time}
Here we  present an application of the results obtained in the previous section. First, 
we assume a given  metric potential, $\phi$ say [c.f. equation \eqref{eq:metric}], and 
then we find the other one  by means of the two equations 
$F_R=c\rho^ne^{bz}$,
and
$F_R=c\rho^ne^{-l^2\rho^2/8}$.
Let us work in  prolate spheroidal coordinates $(x,y)$, with $x \in [1, \infty)$ and 
$y \in [-1, 1]$.
These are related to the 
cylindrical coordinates $(\rho,z)$ through the  relations
\begin{eqnarray}
\label{cyl:coord}
\rho^2=m^2	(x^2 - 1)(1 -y^2)\quad \text{and} \quad z=mxy.
\end{eqnarray}
The line element \eqref{eq:metric} becomes (c.f. equation (4.5.18) in \cite{MC})
\begin{align}
\label{gPB}
\d s^2 = &  - e^{2\phi} \d t^2 + m^2 e^{2(\lambda - \phi)} \left( x^2 - y^2 \right) \left[
\frac{\d
x^2}{x^2 - 1} + \frac{\d y^2}{1 - y^2} \right]\nonumber\\ 
 & + m^2 e^{-2\phi} (x^2-1)(1 - y^2) \d \varphi^2. 
\end{align}

The Einstein vacuum equations in $f(R)$ gravity for a static axially symmetric space-time
can be cast into the form
\begin{subequations}
\begin{eqnarray}
\nabla^2\phi&=&0\\
\lambda_{,x}&=& \tilde\lambda_{,x} + {\beta}{\rho}_{,x} + \Omega z_{,x},\\
\lambda_{,y} &=&\tilde\lambda_{,y} + {\beta}{\rho}_{,y} + \Omega z_{,y},
\end{eqnarray}
\label{odesys}
\end{subequations}
where
\begin{eqnarray}
\label{sch:poten}
\phi=\frac{1}{2}\ln{\left[\frac{x-1}{x+1}\right]} \quad \text{and} \quad
\tilde\lambda=\frac{1}{2}\ln{\left[\frac{x^2-1}{x^2-y^2}\right]}
\end{eqnarray}
are the metric potentials corresponding to the usual Schwarzschild solution in 
standard general relativity. Note that we can relate the prolate coordinates $(x,y)$ to the usual Schwarszchild coordinates $(r,\theta)$ through the relations
(c.f. equation (4.5.19) in \cite{MC})
\begin{align}
 r=m(x + 1) \quad \text{and} \quad \theta= \arccos(y).
\end{align}
This in turn allow us to relate them to \eqref{cyl:coord} by 
\begin{align}
 \rho^2= [(r -m)^2 -  m^2]\sin^2(\theta) \quad \text{and} \quad z=(r - m)\cos(\theta). 
\end{align}
It becomes clear that $\rho=0$ corresponds to $x=1$ and $r=2m$. Thus, in the forthcoming discussion, the reader should be aware that the domain of the prolate coordinates is defined from the horizon up to infinity. Notice as well that
\begin{eqnarray}
{\beta}=\frac{\rho}{2F_R}(F_{R,\rho\rho} - F_{R,zz}) \quad  \text{and} \quad \Omega =\frac{\rho F_{R,\rho z}
}{F_R}.
\end{eqnarray}
Thus, in the same way as the last case, we will obtain a explicit form of
$\lambda$ by considering the different  values of $F_R$.


\subsection{$F_R=c\rho^ne^{bz}$ }

Here we have  
\begin{subequations}
\begin{eqnarray}
\lambda_{,x}&=& \frac{x(1-y^2)}{(x^2-1)(x^2-y^2)}
+ \frac{n(n-1)x}{2(x^2 -1)} \\
&-& \frac{b^2m^2}{2}x(1-y^2) + bmny\nonumber\\
\lambda_{,y} &=&\frac{y}{x^2-y^2} - \frac{n(n-1)y}{2(1-y^2)}\\
&+&\frac{b^2m^2y}{2}(x^2 -1) +bmnx,\nonumber
\end{eqnarray}
\end{subequations}
whose solution is
\begin{eqnarray}
\label{lambda1}
\lambda&=&\tilde\lambda +\frac{n(n-1)}{4}\ln{[(x^2-1)(1-y^2)]}\\
&+& \frac{b^2m^2Q}{4} + bmnxy,\nonumber\\
Q&=&x^2y^2-x^2-y^2.
\end{eqnarray}

Substituting \eqref{sch:poten} and \eqref{lambda1} into the metric \eqref{gPB}, we observe that there are some apparent singularities whenever $x$ and $y$ take the values $\pm 1$.  Note that $x=-1$ is not part of the domain of the prolate coordinate system and that the singularities in $y$ are the trivial pole singularities when $\theta =0$ and $\theta = \pi$. The singularity in $x=1$, however, deserves some deeper analysis. To this end, we compute the two main curvature invariants, i.e. $R=R_a^{\ a}$ and $K=R_{abcd}R^{abcd}$.

Let us assume that $n\geq 1$ and consider the two possibilities, $b=0$ and $b\neq0$, separately.
\begin{enumerate}
\item $b=0$. In this case, the Ricci curvature scalar is
	\beq
	R = -{\frac {n \left( n-1 \right)  \left( {x}^{2}-{y}^{2} \right)  \left( {x}^{2}-{y}^{2}{x}^{2}-1+{y}^{2} \right) ^{-1/2\,n \left( n-1 \right) }}{{m}^{2} \left( x+1 \right) ^{2} \left( -1+{y}^{2} \right)  \left({x}^{2}-1 \right) }},
	\eeq
whereas the Kretchman invariant takes the form
	\beq
	K = -\frac{f(x,y;n)}{2\, \left( x-1 \right) ^{2}{m}^{4} \left( x+1 \right) ^{6} \left( 1-{y}^{2} \right) ^{2}},
	\eeq
where $f(x,y;n)$ is a lesser illuminating expression.  One can easily  show that in the large $x$ limit, both of the above expressions converge to zero. Moreover, inspecting the $n=1$ solution in the `equatorial' plane, $y=0$, one obtains 
	\beq
	R_{n=1} = 0
	\eeq
and
	\beq
	 K_{n=1}=\frac{44}{m^4 (x+1)^6} = \frac{44 m^2}{r^6},
	\eeq
in agreement with the Schwarszchild solution.

\item $b\neq 0$. This case is richer in content. The curvature invariants can be written as
	\beq
	R = \frac{h_1(x,y;n)}{m^2 (x+1)}
	\eeq
and 
	\beq
	K = -\frac{h_2(x,y;n)}{2\, \left( x-1 \right) ^{2}{m}^{4} \left( x+1 \right) ^{6} \left( 1-{y}^{2} \right) ^{2}}.
	\eeq
Just as before, the functions $h_i(x,y;n)$ are long polynomial expressions from which little can be said. Interestingly, a similar analysis of the $n=1$ soution in the $y=0$ plane and taking $b=1$ yields
	\beq
	R_{n=1} = -{\frac {{{\rm e}^{1/2\,{m}^{2}{x}^{2}}}{x}^{2}}{ \left( x+1 \right) ^{2}}}
	\eeq
and
	\beq
	K_{n=1} = \frac{\mathcal{O}(x^7)}{ \left( x+1 \right) ^{6} \left( x-1 \right) {m}^{4}}.
	\eeq
 Here we observe that the regular behaviour of the $n=1$ solution at the `horizon' of the $b=0$ case is lost. What we see here is a true singularity at $x=1$ with no horizon dressing it. 
\end{enumerate}

\begin{figure*}
$$\begin{array}{cc}
\epsfig{width=2.85in,file=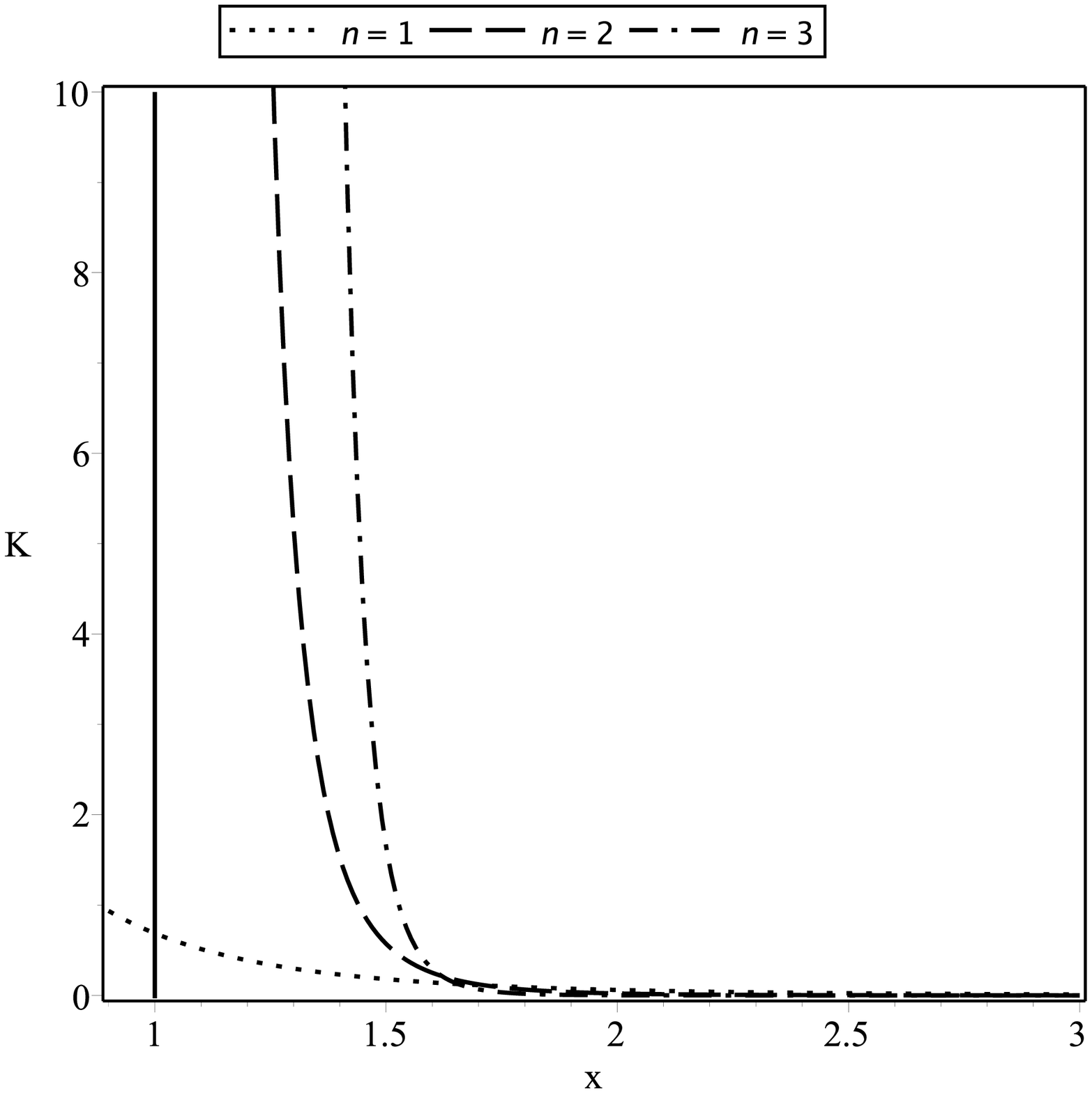} &
\epsfig{width=2.85in,file=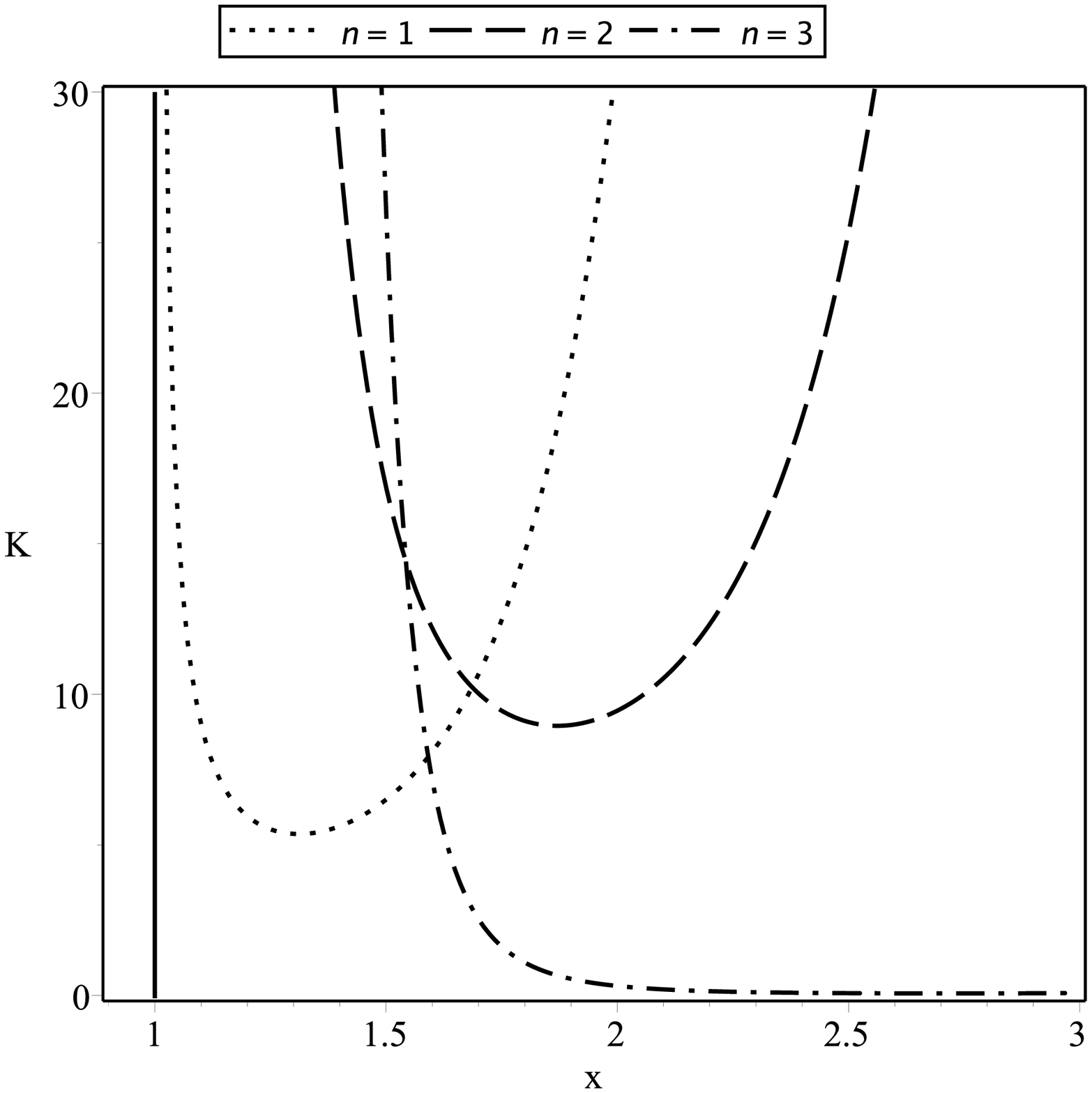} \\
 & \\
 & \\
 & \\
\epsfig{width=2.85in,file=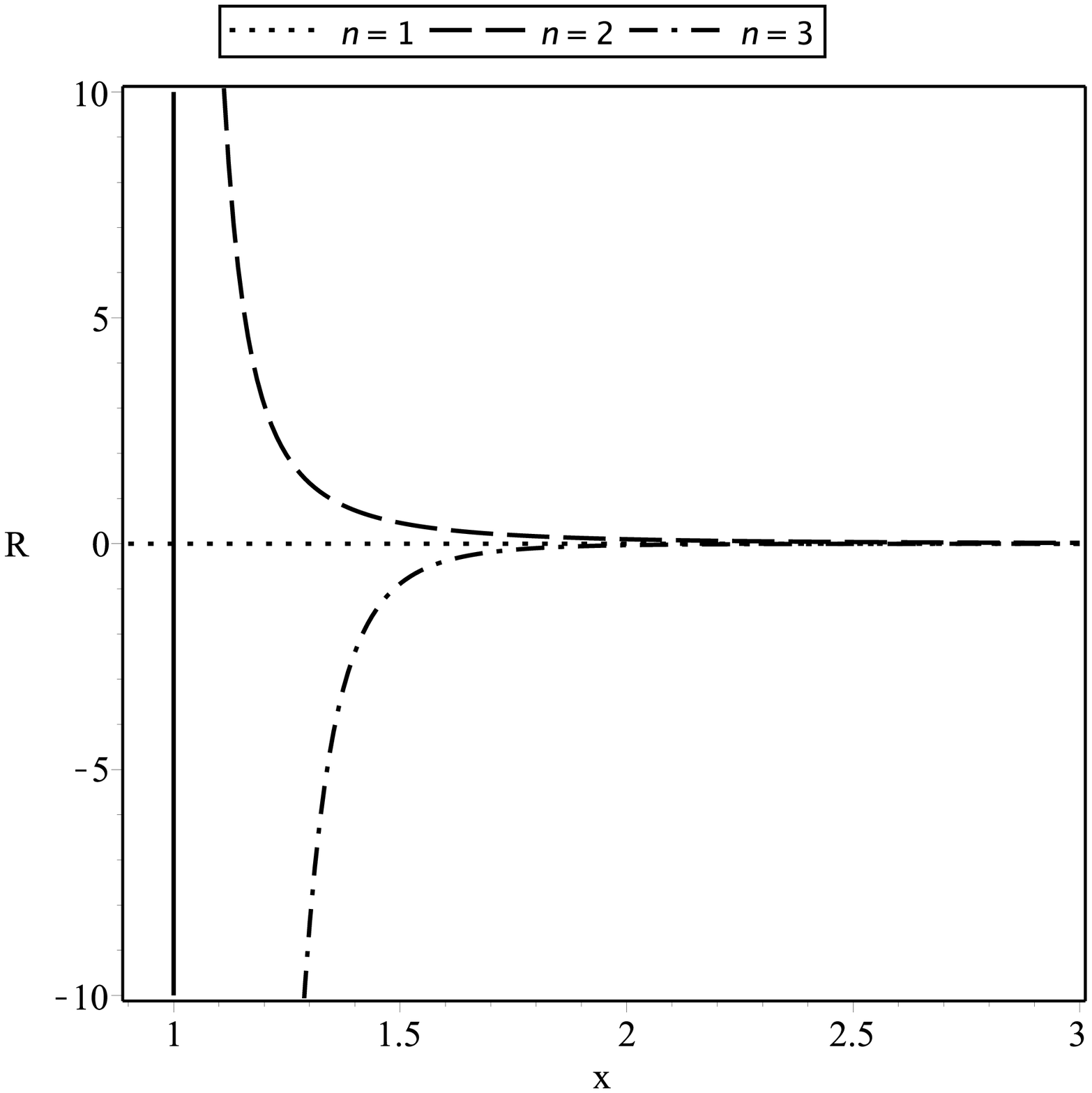} &
\epsfig{width=2.85in,file=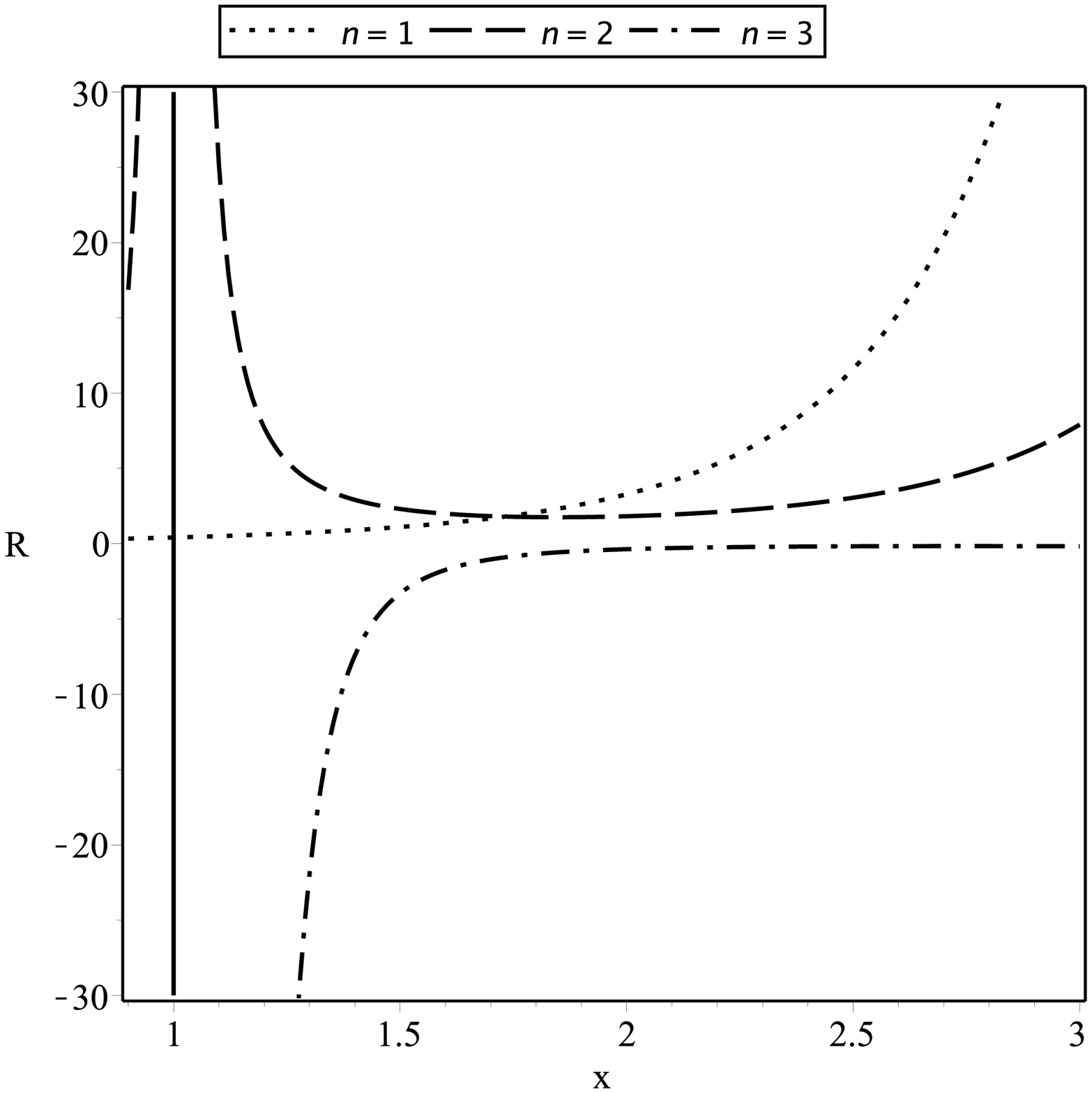} \\
\end{array}$$
\caption{Kretchman  invariant (top plots) and  Ricci curvature scalar (bottom plots)
 as functions of $x$ for  $F_R=c\rho^ne^{bz}$ with  $n=1,2$ and $n=3$ for the values $b=0$ (left plots) and $b=1$ (right plots).}\label{figure1}
\end{figure*}

The behaviour in the vicinity of $x=1$ is shown in figure \ref{figure1} for the soutions with $n=1,2,3$.

\begin{figure*}
$$\begin{array}{cc}
\epsfig{width=2.85in,file=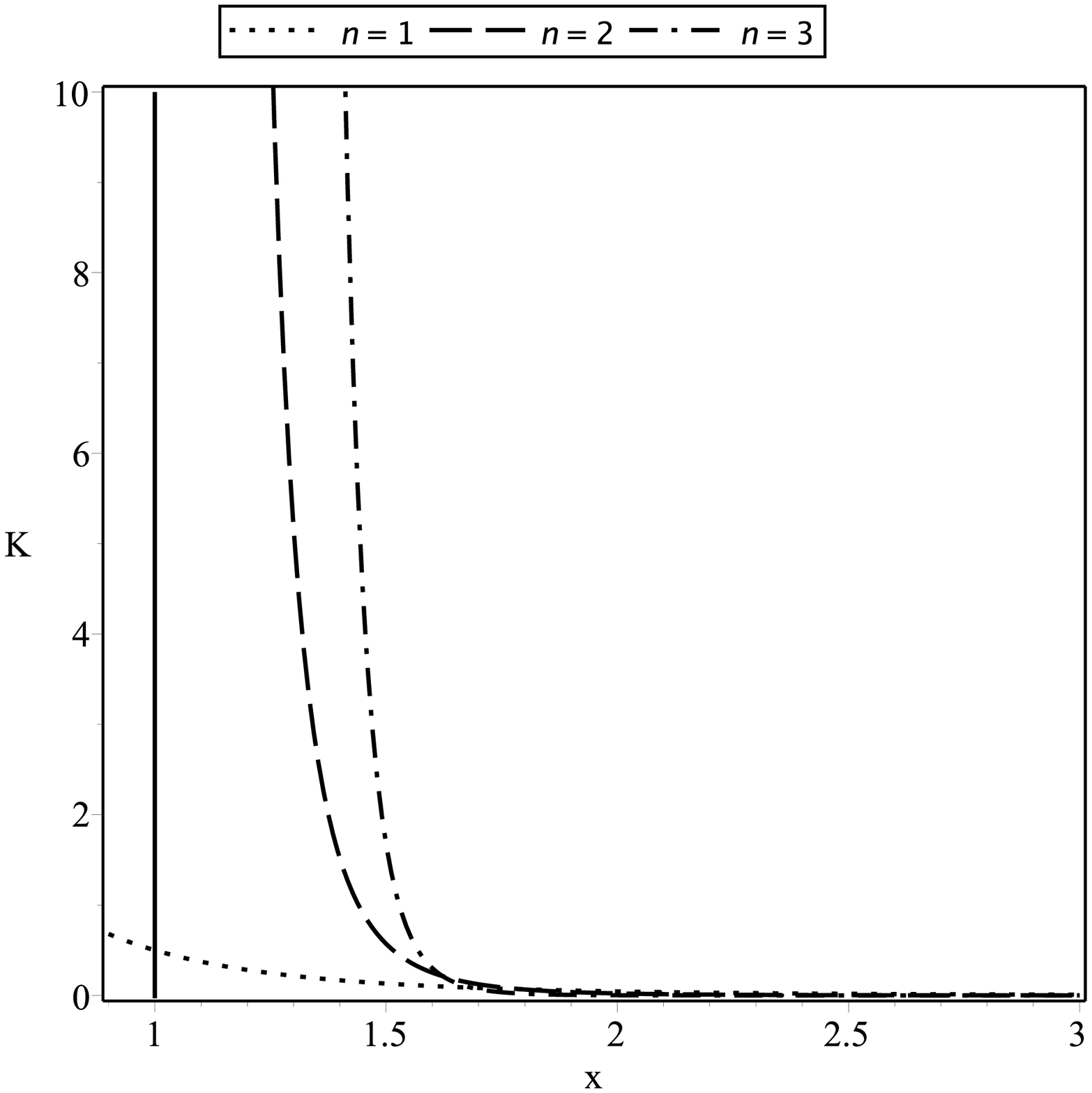} &
\epsfig{width=2.85in,file=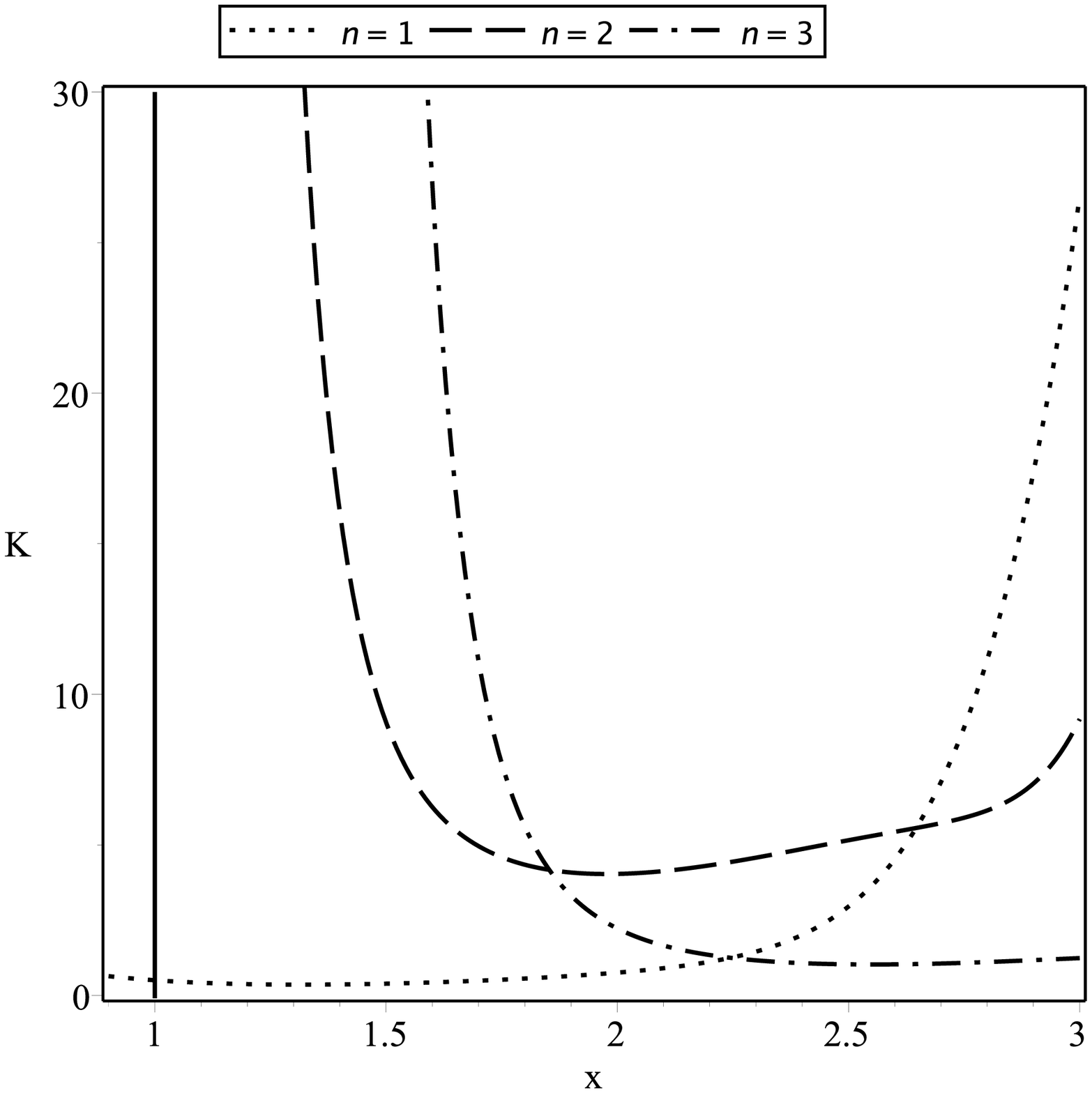} \\
 & \\
 & \\
 & \\
\epsfig{width=2.85in,file=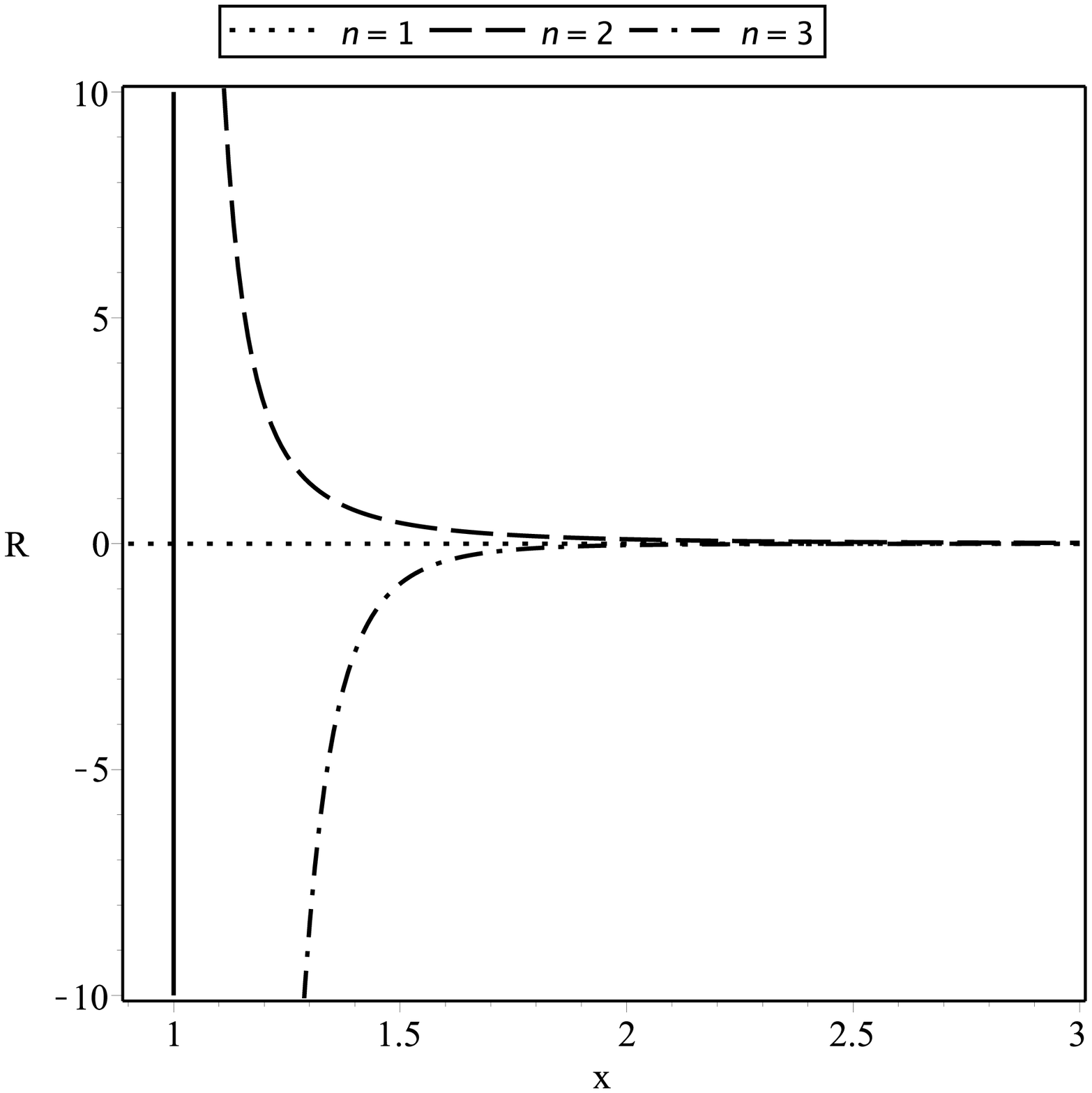} &
\epsfig{width=2.85in,file=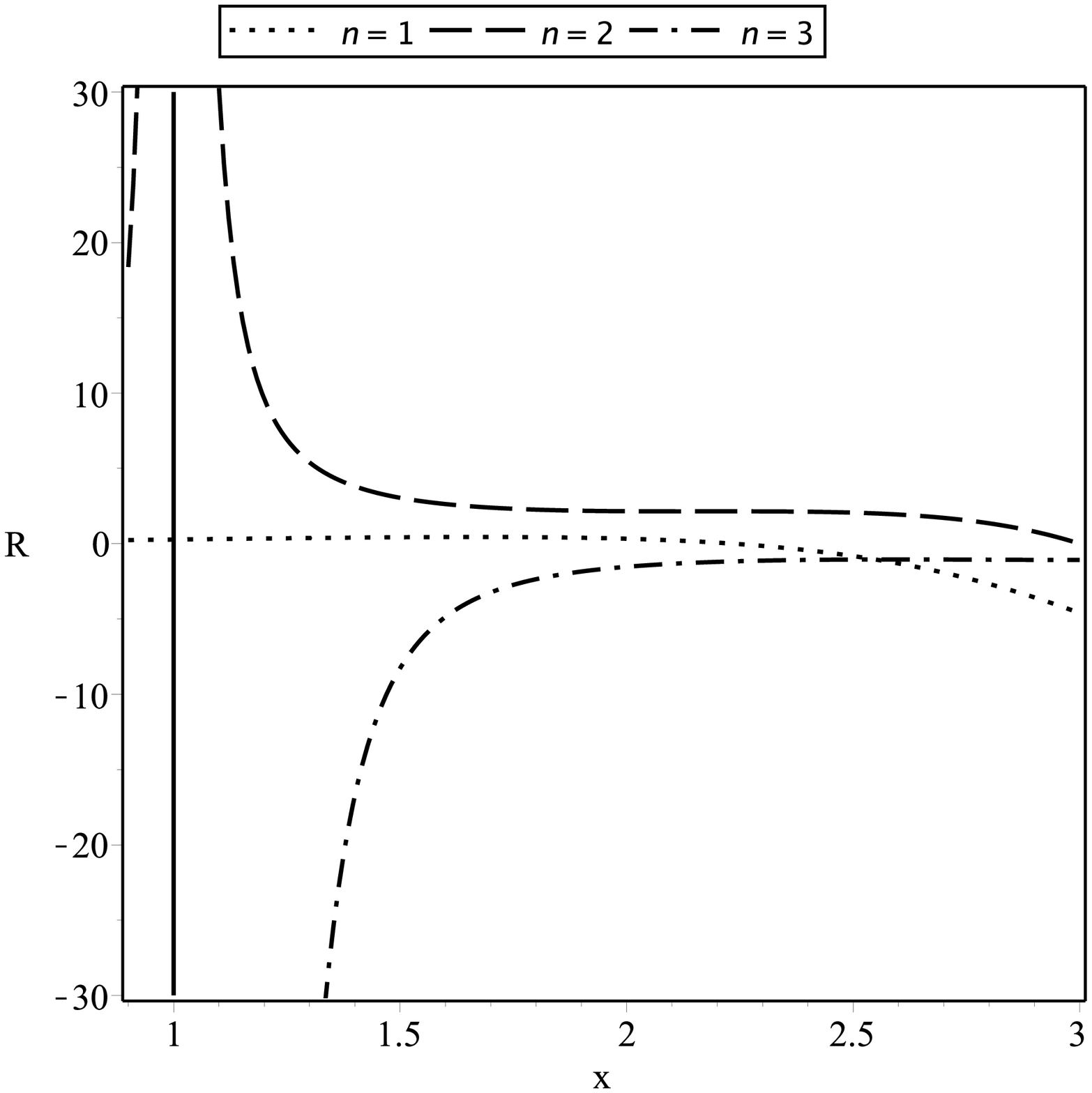} \\
\end{array}$$
\caption{Kretchman  invariant (top plots) and  Ricci curvature scalar (bottom plots)
 as functions of $x$ for  $F_R=c\rho^ne^{-l^2\rho^2/8} $ with  $n=1,2$ and $n=3$ for the
values $l=0$ (left plots) and $l=1$ (right plots).}\label{figure2}
\end{figure*}


\subsection{$F_R=c\rho^ne^{-l^2\rho^2/8}$ }

Now, in this case 
\begin{subequations}
\begin{eqnarray}
\lambda_{,x}&=& \tilde\lambda_{,x} + \frac{x}{32(x^2-1)}P, \\
\lambda_{,y}&=& \tilde\lambda_{,y} - \frac{y}{32(1-y^2)}P,\\
P&=& [16n(n-1) +l^4\rho^4
-4l^2(2n+1)\rho^2]\nonumber,
\end{eqnarray}
\end{subequations}
and the solution is written as
\begin{eqnarray} 
\lambda&=&\tilde\lambda + \frac{n(n-1)}{4}\ln{[(x^2-1)(1-y^2)]}\nonumber\\
&+&\frac{l^2m^2}{128}[l^2m^2(Q+2) + 8(2n +1)]Q.
\label{eq:lambdaschwarzII}
\end{eqnarray}

Just as in the previous case, we analyse the curvature invariants to look for singularities. Again, we split our study into two cases
	\begin{enumerate}
	\item $l = 0$. The curvature invariants are
		\beq
		R = -{\frac {h_3(x,y;n)}{ \left( x+1 \right) {m}^{2}}}
		\eeq
and
		\beq
		K = \frac{h_4(x,y;n)}{2\, \left( x-1 \right) ^{2}{m}^{4} \left( x+1 \right) ^{6} \left( -1+{y}^{2} \right) ^{2}}.
		\eeq
Same as before, the equatorial plane $n=1$ solution reduces to Schwarszchild
	\beq
	R_{n=1} = 0
	\eeq
and
	\beq
	K_{n=1} = 32\,{\frac {1}{ \left( x+1 \right) ^{6}{m}^{4}}}.
	\eeq
	
	\item $l \neq 0$.  This case shares the same singular structure as the $l=0$ case. As can be seen in figure \ref{figure2} for $n=1, 2, 3$. Note that for these soutions, $n=1$ is always regular at the horizon. However, taking $n>1$ always produces a naked singularity at $x=1$.
	\end{enumerate}

Finally, let us note that, in every case, the curvature invariants converge to zero in large $x$ limit.








\section{\label{sec:conclude}Closing remarks}
The issue of  static and axially symmetric solutions in $f(R)$-gravity is a timely topic in the context  of the exact solutions. In this paper, we have
presented an axially symmetric static vacuum solution in Weyl coordinates for $f(R)$
 gravity. In particular,  from the integrability
condition of one of the metric potentials of the Weyl-Lewis-Papapetrou line
element and using the method of separation of variables we have obtained a
general explicit expression for the dependence of $df(R)/dR$ on the $\rho$
and $z$ coordinates and, therefore,  the corresponding general explicit form of $f(R)$. 
Working in prolate spheroidal coordinates, we have  analysed  in detail the `Schwarzschild' solution to the modified field 
equations. We have shown that these particular static and axially symmetric vacuum solutions of $f(R) \neq R$ correspond to naked singularities, as can be seen in the right hand column in figures \ref{figure1} and \ref{figure2}. In particular, one observes that the singularity structure of the case  $F_R = c\rho^n e^{bz}$ is very sensitive to the value of $b$ as can be seen in figure \ref{figure1} where the solution ceases to be  regular at the horizon and becomes  a naked singularity, even in te $n=1$ case.

Finally, it is worth  noting that  the  potentials (\ref{lambda1}) and (\ref{eq:lambdaschwarzII})
were
found by integrating the corresponding system of differential equations [equations \eqref{odesys}]. However, these solutions 
can be obtained directly by using the transformations \eqref{eq:WeylModI} and \eqref{eq:WeylModII}. This result allows us to generate axially symmetric solutions for $f(R)$ from known seeds of the Weyl class of solutions for the Einstein field equations.

This  work is dedicated with great pleasure to Biky (M.V.R.H.) on the occasion of her 23rd
birthday.

\section*{Acknowledgments} 
ACGP was partially supported by a TWAS-CONACYT Postdoctoral
Fellowship Programme. CSLM acknowledges support from CONACYT grant 290679\_UNAM.

\end{document}